\documentclass[10pt,a4paper]{article}
\title{\bf NONLINEAR~DIELECTRIC~RESPONSE OF MICROCOMPOSITE DISPLACIVE FERROELECTRICS.}
\author{Matej Hudak \\ {\it Lab A,} Stierova 23, SK - 040 23  Kosice, Slovak Republic\footnote{Corresponding author}\\hudakm@mail.pvt.sk
	\and Jana Tothova\\ {\it Lab A,} Stierova 23, SK - 040 23  Kosice, Slovak Republic
    \and Ondrej Hudak\\ {\it Lab A,} Stierova 23, SK - 040 23  Kosice, Slovak Republic}
\begin{document}
    \maketitle{}

\section*{PACS Numbers:}

\begin{itemize}
\item 
77.22 Ch Permittivity (dielectric function)
\item 
77.84 f~ Composite materials
\item 
 77.84.-s Dielectric, piezoelectric, and ferroelectric materials
\item 
 42.65.-k Nonlinear optics
\end{itemize}

\section*{Condensed paper title:}

{\it NONLINEAR DIELECTRIC RESPONSE OF MICROCOMPOSITE FERROELECTRICS.}
\newpage
\vspace{1in}
\begin{abstract}
	
A dielectric nonlinear response in model two-phase composites, prepared from
a displacive ferroelectric material with a dominant dielectric response due to
a single oscillator ferroelectric mode and a dielectric material, is characterized
by the existence of new modes due to geometric resonances below the percolation threshold in the ferroelectric matrix. The geometric nature of oscilations
depends on composition and geometric properties of the mixture.
Generation of higher frequency modes is studied within our model and discussed for displacive ferroelectric materials of the $BaTiO_{3}$ type. The Bergman
Representation of the effective dielectric function together with our model for
spectral function recently introduced are used simultaneously with theory of
absorption in finite displacive ferroelectric crystals in quasistatic approximation
to describe the physics of the composite. The boundaries of the present study
are within this quasistatic approximation. New ways how to optimize material
properties for nonlinear optics is outlined.
\end{abstract}

\newpage
\tableofcontents

\newpage
\section{Introduction.}

Dielectric response of ferroelectric microcomposite materials formed from order-disorder ferroelectric and dielectric grains as components have been shown recently [1] to display features not present in pure order-disorder ferroelectrics
nor in pure dielectrics. Finite size effects of grains (depolarization effects and
intergrain interactions) may lead to new effective properties of the resulting
medium. Different effective medium theories (Maxwell-Garnet Theory, the Effective Medium Approximation theory and the Bergman Representation of the effective dielectric function) were used in [1] together with the theory of absorption in finite ferroelectric crystals in quasistatic approximation to describe the physics of the composite. One of main results concerned existence of a relaxation mode due to geometric resonances which lead to new absorption peak below the
percolation threshold in the ferroelectric matrix. The soft mode still exists in
this matrix too, with renormalized mode strength and unchanged relaxation
frequency. Another feature found is hardening of the soft mode in the dielectric phase in which finite ferroelectric clusters exist. High-frequency dielectric function is also modified in a nonlinear way as concerning the concentration dependence of ferroelectric particles.
It is the aim of this paper to study dielectric response in ferroelectric microcomposites in which the ferroelectric component is of the displacive type instead of the order-disorder type, and the dominant response of which to electric field is due to an oscillator type polar mode. Thus this our study represent extension of our previous one concerning order-disorder type materials [1], to a more general class of materials with oscillator type of mode. Note that the
order-disorder type characteristics are limiting case of the oscillator type for
larger values of the damping constant. A still more realistic model should be
based on the multimode approach, however, the single mode case, as adopted
here, gives main qualitative features of the new characteristics present in the
microcomposite. It corresponds to the multimode case in which the soft mode is
sufficiently softened below other resonance modes in the lower frequency region.
All but the soft mode are semiempirically taken into account in our approach
via the high-frequency dielectric constant.
To clarify which relaxation processes are present in such composites we use
the Bergman Representation for the effective dielectric function together with
our model for the spectral function introduced in [1]. This choice is supported
by the fact, that the mentioned representation covers the whole spectrum of the
particle concentrations, reflects existence of the percolation transition and that
our model spectral function gives in the limiting cases of small concentrations
results which are identical in the main order with those obtained by using the
Maxwell-Garnet Theory. It is well known that the Maxwell-Garnet Theory is
appropriate and sufficiently simple to identify qualitative changes of the 
composite properties at small concentration limits of one of the components only.
The Bergman Representation is useful for discussions of the dynamic properties
of the effective dielectric function, which go beyond results obtained within the
frame of the other known theories. First the model of two-phase microcomposite ferroelectric medium is introduced in which the ferroelectric part consists of the displacive material. We
start with some results following from theory of absorption in finite displacive
ferroelectric crystals in quasistatic approximation.
One of the main results of this paper consists of confirmation of presence
of the new relaxation mode due to geometric resonance also in displacive type
ferroelectric microcomposites. Another feature discussed in this paper is 
hardening of the soft mode in the phase with infinite dielectric matrix in which finite
ferroelectric clusters exist. Below the percolation threshold the soft mode exists
with renormalized mode strength.
High-frequency dielectric function is also modified resulting in a nonlinear
dependence on concentration of ferroelectric particles. Spontaneous birefringence of ferroelectric materials usually depends linearly or
quadratically in spontaneous polarization depending on the character of the
paraelectric phase. In the last section a nonlinear susceptibility of microcomposite  displacive ferroelectrics is calculated.

\section{Model of two-phase microcomposite displacive \\ ferroelectric medium.}

Model microcomposite displacive ferroelectric materials are considered here to
be those materials which are composed from two phases: a normal dielectric material, and a ferroelectric material undergoing in bulk at some temperature a second order displacive phase transition from paraphase to ferroelectric phase. For
simplicity we assume that our model microcomposite consists of packed
spherical particles. We assume that the interparticle free volume due to random
packing of spheres is semiempirically taken into account in our calculations of 
effective dielectric properties of the composite due to properties of the Bergmann
Representation spectral function, see in [1], [2] and [3].
Let our composite be based on two basic materials. One of them is a displacive ferroelectric material (F). Let its relative volume concentration be $(1-x)$,
where $x$ is the volume concentration of dielectric grains: $0 \leq x \leq 1$. Let us
further assume that the soft mode behavior of the bulk is conserved in the material from which spherical ferroelectric particles are formed, we assume that these particles are sufficiently large. There are, however, corresponding size and shape modification of the dielectric response of such grains, the soft mode behavior of particles is assumption which is based on corresponding results of our recent calculations [4] concerning dynamic dielectric response of small ferroelectric particles.
A contribution to the total effective permittivity due to contribution $\varepsilon_{F}$ of
the ferroelectric F grains is at lower frequencies in good approximation described
by a single oscillator process characteristic for displacive type systems:
\begin{equation}\label{1}
\varepsilon =  \bigtriangleup \varepsilon + \varepsilon_{\infty}
\end{equation}
where:
\[\bigtriangleup \varepsilon = \frac{f}{\omega_{0}^{2} - \omega^{2} + i . \omega . \Gamma },
\]
$\varepsilon_{\infty}$ is high-frequency dispersions mechanisms contribution, $\Gamma$ is the inverse relaxation time, $\omega$ is frequency and $\omega_{0}$ is the soft mode resonance frequency. Note
that in the case of order-disorder ferroelectrics one correspondingly considers
the limit of an overdamped oscillator.
We assume that the ferroelectric (F) material is characterized in bulk by a
single second order phase transition critical temperature $T_{C}.$ Contributions of
other hard modes present in the bulk material to the dynamic dielectric function
$\varepsilon_{F}$ of the pure ferroelectrics F are included approximately in the high-frequency
contribution $\varepsilon_{\infty}$. We consider those lower frequencies at which the soft mode
contribution dominates the response. The oscillator strength $f$, the soft mode
frequency $\omega_{0}$ and the inverse relaxation time $\Gamma$ in this material are temperature
dependent quantities in general. The soft mode is softening at the center of the
Brillouin zone:
\begin{equation}\label{2}
\omega_{0} = \omega_{0,HT} . a^{-1}. \mid 1 - \frac{T_{C}}{T} \mid.
\end{equation}
The numerical constant $a$ takes the following values: $a = 1$ above the 
transition temperature $T_{C}$, and $a = \frac{1}{2}$ below that temperature, in correspondence with the Landau theory [5] of such phase transitions. It is, however, convenient to
neglect weak temperature dependence of the oscillator strength $f$ and of the
inverse relaxation time $Γ$ and to consider both quantities as material constants.
This assumption is in general not correct: there is certainly some temperature
dependence of the single-atomic relaxation modes. We assume that this assumption is valid
near the transition temperature.
As concerning particle diameter $d $ a distribution of diameters should
exist there in real systems with some average diameter and with some characteristic variance of diameters. In correspondence with existing methods of grains
preparation we again assume that the variance of diameters is negligible and that
its value in our model composite is the same for the whole range of concentrations $x$. Let us note that for very large diameters $d$ the paraelectric-ferroelectric phase transition may be well described by the formula (2) in which neither critical temperature $T_{C}$ nor material constants are size dependent. For smaller values
of the diameter $d$ the critical temperature $T_{C}$ and the material constants are quantities  dependent on the diameter $d$, as it was recently discussed in [4]. 
Moreover, we assume that there is no dispersion of the transition temperatures
in ferroelectric grains.
The other basic material from which our ferroelectric composite medium  is composed is a dielectric material (D). We assume that its permittivity is in
considered temperature and frequency region in good approximation described
by a constant real dielectric function $\varepsilon_{D}:$
\begin{equation}\label{3}
\varepsilon_{D} \approx const.
\end{equation}
The volume concentration of macroscopic domains, formed from grains 
containing the dielectric D is $x$.
To perform orientational numerical estimates one may use such numerical values
of material constants occurring in our model system, which are characteristic [6] 
for displacive ferroelectric materials of the $BaTiO_{3}$ type: $T_{C} = 391~K$, the Curie
constant $C = 1.7~10^{5}~K$, the high temperature inverse time $\frac{\Gamma}{\omega_{0,HT}^{2}.a^{-1}} = 10^{13}~s,$  and $\varepsilon_{\infty} = 5.$
Note that the Curie constant $C$ is related to the oscillator strength
via the relation $C = \frac{f}{\omega_{0,HT}^{2}.a^{-1}}.$
The high temperature limiting frequency $\omega_{0,HT}$ need not to be specified if the frequency $\omega$ is measured in units of $\omega_{0,HT}.$  The permitivity of the dielectric grains is considered to be the same as in the bulk dielectric  material, which gives a typical value $\varepsilon_{D} = 10$.

A starting point in our calculations is to consider dielectric response of a finite-size ferroelectric grain in a dielectric medium. Differences in dielectric
functions between two materials, finite size and depolarization field effects
lead to a dielectric response which is different from that of a bulk ferroelectric
material. When concentration of grains increases then the electric field which
is external to a given grain is modified due to other grains. Thus multiplication
effects due to mutual influence of many grains in a given medium may lead to
new phenomena as concerning the dielectric response of such a system. The
general approach to the description of the effective medium is based on the
Bergmann Representation of the effective dielectric constant. This approach is
used in our last section but one using a simple model spectral density function,
the limiting cases of which lead to the Maxwell-Garnett Theory and which
simultaneously describes well the percolation transition in a very transparent
form. 

\section{Theory of absorption in finite displacive type \\	ferroelectric crystals in quasistatic approximation.}

A ferroelectric displacive type ellipsoidal sample inserted in a medium with
dielectric constant $\varepsilon_{D}$ responses to a homogeneous external electric field ${\bf E}_{M}.$ 
A resulting electric field $\bf E$ in the quasistatic approximation is homogeneous too,
and is equal to, [7]:
\begin{equation}\label{4}
{\bf E} = \frac{\varepsilon_{D}. {\bf E}_{M} - \gamma. {\bf D}}{\varepsilon_{D}.(1-\gamma)},
\end{equation}
where the depolarization field is given by:
\[ {\bf E}_{dep} = 4.\pi.\gamma . \frac{\varepsilon - \varepsilon_{D}}{\varepsilon_{D}.(\varepsilon - 1)}. {\bf P}. \]
Here $\varepsilon$ is a dielectric function of the sample, $\gamma$ is a depolarization factor of the grain for which $0 \leq \gamma \leq 1$. Note that spherical shape of isotropic grains
represents the most simplified version of grains shape description. It will be
used here, thus the depolarization factor is $\gamma = \frac{1}{3}.$
It is convenient to introduce the effective dielectric constant $\varepsilon^{*} $ 
as a change of the electric induction field due to a change of the electric field in zero external field ${\bf E}_{M}$ instead of a dielectric constant $\varepsilon$ as change of the electric induction field due to the change of the electric field {\bf E} in zero field limit. Then we obtain:
\begin{equation}\label{5}
\varepsilon^{*} =   \frac{\varepsilon . \varepsilon_{D}}{\varepsilon_{D}. (1-\gamma)+ \gamma . \varepsilon}
\end{equation}
After substitution of the soft mode behavior for the ferroelectric material (1) we
obtain:
\begin{equation}\label{6}
\varepsilon^{*} = \bigtriangleup \varepsilon^{*} + \varepsilon^{*}_{\infty}.
\end{equation}
where
\[ \bigtriangleup \varepsilon^{*} = \frac{f^{*}}{\omega^{*2}_{0} - \omega^{2} + i. \omega . \Gamma}. \]
The renormalized soft mode strength $f^{*}$ is given by:
\begin{equation}\label{7}
f^{*} = f . \frac{\varepsilon^{2}_{D}.(1-\gamma)}{[\varepsilon_{D}.(1-\gamma)+\gamma. \varepsilon_{\infty}]^{2}},
\end{equation}
and the renormalized (former soft, now hard) mode frequency behavior is:
\begin{equation}\label{8}
\omega^{*2}_{0} = \omega^{2}_{0} + \frac{f. \gamma}{\varepsilon_{M}.(1-\gamma) + \gamma . \varepsilon_{\infty}}.
\end{equation}
The inverse relaxation time $\Gamma$ is found to be unchanged in this grain. The
frequency behavior of $\varepsilon^{*}$ is qualitatively the same as that of $\varepsilon$. Quantitatively both quantities are not the same, the oscillator strength is decreased 
$( f^{*} < f ) , $ and the mode frequency is substantially increased $(\omega^{*2}_{0} >> \omega^{2}_{0}.$ Moreover the high-frequency dispersions mechanisms contribution $\varepsilon_{\infty}$ changes to $\varepsilon^{*}_{\infty}:$
\begin{equation}\label{9}
\varepsilon^{*}_{\infty} = \frac{\varepsilon_{\infty}.\varepsilon_{D}}{\varepsilon_{D}.(1-\gamma)+\gamma.
\varepsilon_{\infty}}.
\end{equation}
The high frequency contributions are also renormalized and the high-frequency
dielectric function is smaller for dielectric medium with dielectric constant smaller
than the original high-frequency constant of the pure ferroelectric material, and
vice versa. Results of this section concern the single ferroelectric crystallite in a 
dielectric homogeneous medium. The composite material corresponds with this situation in the limit of very small concentration of ferroelectric crystallites with very
large intergrain distances. Increasing concentration of crystallites their distance
decreases and the electric field acting on a given crystallite becomes modified due
to presence of other crystallites. Consequently the effective dielectric response
changes from that described above. To describe corresponding modifications we
have to include not only effects of a finite crystallite dimensions but also a finite
intergrain distances effects via concentration effects. We consider such effects
in next sections. To discuss the whole concentration range we need to consider
not only the situation in which dielectric medium contains ferroelectric particles
but also the inverted one in which ferroelectric matrix contains small dielectric
particles.

\section{Bergmann Representation of the effective dielectric function.}

A two-phase composite with sharp boundaries between two phases has been shown
[3] to be represented in general (in the quasistatic approximation) as a sum
of simple poles. The weight of each pole is given by the specific spectral density
function characterizing the geometry of a given composite and of a topology of
clusters formed from particles of a given type. Such a representation is known
as the Bergman Representation [3]. Introducing abbreviation $t \equiv \frac{\varepsilon_{D}}{\varepsilon_{D} - \varepsilon}$
one finds that the effective dielectric function of the composite has in the Bergman Representation the form:
\begin{equation}\label{10}
\varepsilon_{eff} = \varepsilon_{D} . (1 - (1-\gamma)\int_{0}^{1} \frac{g(n, x)}{t - n}.dn).
\end{equation}
Here $g(n, x)$ is the spectral density function containing all information about
geometry and other properties of the composite. It is possible to show, using formulas from [3], that the zeroth and the first order moments of the spectral density function $g(n, x)$ are given by:
\begin{equation}\label{11}
\int_{0}^{1} g(n, x).dn = 1
\end{equation}
which is a normalization condition, and by:
\begin{equation}\label{12}
\int_{0}^{1} n.g(n, x) . dn = \frac{x}{3}.
\end{equation}
The zeroth moment is found to hold for general conditions, but the first moment
has the form (12) only in the case of statistical isotropy [3], which is assumed
to be present in our composite further in our paper. Using our model parametrization of the
permittivity of dielectric particles and of the ferroelectric particles, see (1), the
effective permittivity in the Bergman Representation takes the form:
\begin{equation}\label{13}
\varepsilon_{eff} = \varepsilon_{D} - \varepsilon_{D}.(1-x). \int_{0}^{1} 
\frac{\varepsilon_{D} - \varepsilon_{\infty}}{\varepsilon_{D}.(1 - n) + \varepsilon_{\infty}.n}.
g(n, x).dn +
\end{equation}
\[    +    \varepsilon_{D}.(1-x). \int_{0}^{1} 
\frac{\Phi^{*}(n)}{\omega^{*2}_{0} - \omega^{2} + i. \omega. \Gamma}.
g(n, x).dn.        \]
The first two terms describe the modified dielectric high-frequency response
and the last term represents a sum of contributions due to oscillator processes
generated by the microcomposite, each process corresponding to a specific value
of the parameter $n$. It contributes with the modified strength $\Phi^{*}(n),$
the weight $g(n, x)$, and the resonance frequency $\omega_{0}^{*}(n),$
where:
\begin{equation}\label{14}
\Phi^{*}(n) = \frac{f.\varepsilon_{D}}{(\varepsilon_{D}(1-n)+ \varepsilon_{\infty}.n)}
\end{equation}
and where:
\begin{equation}\label{15}
\omega^{*2}_{0}(n) = \omega_{0}^{2} + \frac{f.n}{\varepsilon_{D}.(1-n)+\varepsilon_{\infty}.n}.
\end{equation}
Dependence on the geometric and topological factors is fully contained in the
spectral density function $g(n,x)$. From (14) and (15) we see that the $n=0$
mode contributes with the same frequency as the original ferroelectric mode,
its strength is weighted by $(1-x).g(0, x)$.
Parameterization of the spectral density, corresponding to our model ferroelectric composite will be in our paper based on the same assumptions as in our recent paper [1].
There are in general two main contributions in composites to the spectral
function, [2], a peak between $n=0.2$ and $n=0.3$ assigned to geometrical resonances of finite clusters and isolated particles, and a steep delta-like contribution at $n=0$ assigned to the resonances of an infinite percolating cluster. 
As a simplification we will demand in our paper that at small concentrations of
dielectric particles the Maxwell-Garnett Approach approximation should result
and at large concentrations $(x \approx 1)$ the Maxwell-Garnett Theory approximation
should be valid for description of a dielectric matrix containing ferroelectric particles. These our assumptions about the spectral function can be easily satisfied
assuming the following two-peak form of the spectral density [1]:
\begin{equation}\label{16}
g(n, x) = g_{0}(x). \delta(n) + a(x). \delta(n - n_{T}(x)).
\end{equation}
In (16) the first delta peak is localized at $n=0$, in correspondence with general
expectations [2], and $g_{0}(x)$ describes its relative intensity for a given composite
concentration x. This peak is due to an infinite ferroelectric cluster contribution,
if this type of cluster is present in the composite. The second delta peak in (16) is
localized at some nonzero value of the parameter $n = n_{T}(x)$, here $a(x)$ describes
the relative intensity of this peak. While the second peak is expected on the
general grounds to be of the wide shape, here in our model its shape is very
narrow. The localization of the second peak is here assumed to be dependent
on the concentration x of the dielectric particles:
\begin{equation}\label{17}
n_{T}(x) = \frac{x}{3}
\end{equation}
for x above the percolation threshold, which was already determined in the previous section at the critical concentration $x_{C} = \frac{2}{3}.$ In this region an infinite dielectric
cluster is present and there are finite ferroelectric and other dielectric clusters.
The concentration dependence of this contribution due to geometric resonances
reflects the Maxwell-Garnett Theory type of the effective medium behavior of
finite clusters forming some kind of a dipolar medium. Below the percolation
concentration we assume that the geometric resonances due to remaining finite
ferroelectric and dielectric clusters are present simultaneously with infinite 
ferroelectric cluster. This region is characterized by a concentration independent
position $n_{T} = \frac{2}{9}$  of the corresponding peak in the spectral density function. For concentrations below the critical concentration both relative intensities
have the following form [1]:
\begin{equation}\label{18}
g_{0}(x) = \frac{x_{C} - x}{x_{C}},
\end{equation}
\[  a(x) = \frac{x}{x_{C}}.  \]
In this region the weight $a(x)$ decreases from the value $1$ (for $x = 0$)to the value $0$ (for $x = x_{C}$). The weight $g_{0}(x)$ decreases here from the value $1$ (for $x = 0$) to the value $0$ (for $x = x_{C}$). Note that the sum of weights is one: $ a(x) + g_{0}(x) =1$. Above the percolation transition we assume the constant concentration independent form of the densities $g_{0}(x) = 0$ and $a(x) = 1$. Note that the spectral density function (16) satisfies the zeroth moment and first moment conditions in the whole range of concentrations $x$. One should note that a broader peak in the spectral density function would lead to many other oscillator-like resonance terms in the effective dielectric function
describing a distribution of geometric resonance frequencies. Such a picture
would be more acceptable on the intuitive base. Here we consider a simple case
of a single (effective) geometric resonance. Our model approach, however, still
contains essential features of the realistic response function and represents some
extension of the theory behind the Maxwell-Garnet Theory as well as behind the
Effective Medium Approach. In this way we succeeded to formulate physically
more acceptable model response function for the ferroelectric composite which
is still sufficiently clear in its structure. Substituting our spectral density function from (16) to the expression for the effective dielectric response function (10) we obtain:
\begin{equation}\label{19}
\varepsilon_{eff} = \varepsilon_{D} - (1-x). g_{0}(x)(\varepsilon_{D}-\varepsilon{\infty})-
\end{equation}
\[ - (1-x).\varepsilon_{D}.a(x). \frac{\varepsilon_{D}-\varepsilon{\infty}}{\varepsilon_{D}.(1-n_{T})+\varepsilon_{\infty}.n_{T}}+  \]
\[ + (1-x).g_{0}(x).\frac{\varepsilon_{0}-\varepsilon_{\infty}}{\omega^{2}_{0}-\omega^{2}+ i. \omega.\Gamma} + \]
\[ + \varepsilon_{D} . (1-x).a(x). \frac{g^{*}(n_{T})}{\omega^{*2}_{0}(n_{T})-\omega^{2}+ i. \omega.\Gamma}.  \]
The first three terms describe modified high-frequency response, the last two
terms describe frequency response in the form of two simple resonances. The last
but one term corresponds to the response of an infinite ferroelectric cluster below the
percolation threshold. The ferroelectric displacive-type infinite cluster responses
in the same way as it is assumed to be present in pure bulk ferroelectric material.
However, its relative intensity is decreased by the factor $(1 - x)g_{0}(x)$ and vanishes
at and above the percolation threshold. The last term describes a response of
small finite clusters and isolated particles in the whole concentration range.
This response process vanishes at concentration $x=0$, where a pure ferroelectric
matrix is present, and at concentration $x=1$, where a pure dielectric matrix is
present. In both limits there are present no additional contributions due to
dipolar properties of the medium. The strength of finite clusters response is
given by $\varepsilon_{D}.(1 - x).a(x)\Phi^{*}(n_{T}).$
Their frequency is in the form similar to that of a finite particle from (8), in which however instead of the depolarization factor $\gamma$ the factor $n_{T}$ is substituted. It is easy to show that the new resonance frequency is always hard, it is not corresponding to the soft mode. At the ferroelectric phase transition temperature this relaxation frequency remains nonzero,  it does not completely softens. The relative strength of both relaxation processes below the percolation
transition changes. There may exist, however, region in space of physical 
parameters (concentration, high-frequency permittivities, Curie constant etc.) in
which both contributions become comparable.

\section{Nonlinear susceptibility of microcomposite \\ displacive ferroelectrics.}

Spontaneous birefringence of ferroelectric materials usually depends linearly or
quadratically on spontaneous polarization depending on the character of the
paraelectric phase [8]. We consider here the $BaTiO_{3}$ type of materials, for
which the paraelectric phase is centrosymmetric and thus the quadratic dependence occurs. Nonlinear coefficients $d_{ijl}$ in this material are linearly dependent
on the spontaneous polarization. Nonlinear polarization ${\bf P}_{NL}$ is defined by:
\begin{equation}\label{20}
P_{NL,i}(\omega_{1} \pm \omega_{2}) = \sum_{jk} d_{ijk}.E_{j}(\omega_{1}).E_{k}(\omega_{2}).
\end{equation}
Here $d_{ijk}(2. \omega)$ are coefficients ($i, j, k, .. $ denote axises) at doubled frequency $\omega$ and are defined by:
\begin{equation}\label{21}
d_{ijk} = \sum_{lmn} \chi^{L}_{il}(2.\omega). \chi^{L}_{jm}(\omega).\chi^{L}_{kn}(\omega). \delta_{lmn}(2.\omega).
\end{equation}
Here $L$ is the index for the linear susceptibility. The sign $+$ or $-$ in the
relation (21) above depends on the character of the process. As in [8] dispersion of
the coefficients $\delta_{lmn}$ is assumed to be small.

Assuming an isotropic approximation for the composite susceptibility (which is related to the the effective dielectric response function described above) we obtain:
\begin{equation}\label{22}
\chi^{L}_{il} = \delta_{il} \chi_{comp},
\end{equation}
where $\chi_{comp}$ is the composite dielectric susceptibility in an isotropic medium 
approximation, and $\delta_{il}$ is the Kronecker delta. Studying the longitudinal polarization, we obtain for the nonlinear polarization $P_{NL,i}:$
\begin{equation}\label{23}
P_{NL,i}(\omega_{1} \pm \omega_{2}) = \sum_{ijk}.d_{ijk} E_{j}(\omega_{1}).E_{k}(\omega_{2}),
\end{equation}
where here $d_{ijk}$ coefficients for $\omega_{1} = \omega_{2} = \omega$ is $d_{ijk}(2.\omega):$
\begin{equation}\label{24}
d_{ijk}(2.\omega) = \chi^{2}_{comp}(\omega). \chi_{comp}(2. \omega) \delta_{ijk}(2.\omega).
\end{equation}
Here $(comp)$ is the index for the composite susceptibility.
Second harmonic generation is thus described by nonlinear longitudinal $(l)$ polarization $P_{NL,l}$ as:
\begin{equation}\label{25}
P_{NL,l} = \delta_{lll} E_{l}^{2}(\omega) . \chi^{2}_{comp}(\omega. \chi_{comp}(2.\omega).
\end{equation}
The strength of the new geometric resonance pole and the frequency of the new geometric resonance pole are both temperature and composition dependent quantities. Together with another free parameter, the value of the dielectric constant of the dielectric component, a
wide range of microcomposite properties may be prepared depending on the
targeted second generation harmonics. 

\section{Discussion.}

In sections above, Introduction, Model of two-phase microcomposite displacive ferroelectric medium, Theory of absorption in finite displacive type ferroelectric crystals in quasistatic approximation, Bergmann Representation of the effective dielectric function and Nonlinear susceptibility of microcomposite  displacive ferroelectrics we discussed theory of absorption in finite displacive type ferroelectric crystals in quasistatic approximation two-phase microcomposite displacive ferroelectric medium using the Bergmann representation of the effective dielectric function and nonlinear susceptibility of this microcomposite  was calculated.

Let us assume that the above described nonlinear coefficients $d_{ijk}$ are macroscopic quantities averaged over distances much larger than the average fluctuation length of the microcomposite. It is clear that the second harmonic generation in microcomposites due to geometric resonances will display new interesting properties. Changing pure composition concentration, changing chemical composition, eventually introducing the third composite component (theory of which is a little bit more complicated than the theory outlined above) one can obtain a wide diapason of composite properties as concerning the electromagnetic field response. For example optimization of materials with electrooptic properties expected for given device application may be enhanced using nonlinear properties of microcomposites described above.

\section*{Acknowledgment}

One of authors (O.H.) would like to express his sincere thanks to Dr. V. Dvorak and Dr. J. Petzelt from Institute of Physics, AS CR Prague (discussions on displacive ferroelectrics), to Dr. J. Gavilano for his discussions during his visit giving seminar in the PSI Villigen (the visit was financially supported by PSI Villigen), and also to Prof. F. Cordero and Dr. F. Craciun from Instituto dei Sistemi Complessi Roma for their discussions on displacive ferroelectrics.

\end{document}